\documentclass[preprint]{aastex}
\usepackage[T1]{fontenc}
\usepackage[latin1]{inputenc}
\usepackage{graphics}
\newcommand{\mla}{\la}            
\newcommand{\mga}{\ga}            

\begin{document}

\title{Implications of the $\gamma$-ray Polarization of GRB 021206}

\author{}
\author{Ehud Nakar\altaffilmark{1}, Tsvi Piran\altaffilmark{1}
and  Eli Waxman\altaffilmark{2,3}} \altaffiltext{1}{Racah
Institute for Physics, The Hebrew University, Jerusalem 91904,
Israel; udini@phys.huji.ac.il, tsvi@phys.huji.ac.il}
\altaffiltext{2}{Department of Condensed Matter Physics, Weizmann
Institute, Rehovot 76100, Israel; waxman@wicc.weizmann.ac.il}
 \altaffiltext{3}{Incumbent
of the Beracha foundation career development chair}

\begin{abstract}
We compare two possible scenarios for the producing of high level
of polarization within the prompt emission of a GRB: synchrotron
emission from a relativistic jet with a uniform (in space and
time) magnetic field and synchrotron emission from a jet with a
random magnetic field in the plane of the shock. Somewhat
surprisingly we find that both scenarios can produce a comparable
level of polarization ($\sim 45-50$\% for the uniform field and
$\sim 30-35$\% for a random field). Uniform time independent
field most naturally arises by expansion of the field from the
compact object. It requires a $10^{12}$G field at the source and
a transport of the field as $\propto R^{-1}$.  It {\it does not}
imply Poynting flux domination of the energy of the wind. There is
a serious difficulty however, within this scenario, accounting for
particle acceleration (which requires random magnetic fields) both
for Poynting flux and non-Poynting flux domination. Significant
polarization can also arise from a random field  provided that
the observer is located within $1/\Gamma$ orientation from a
narrow ($\theta_j \sim 1/\Gamma$) jet. While most jets are wider,
the jet of GRB 021206 from which strong polarization was recently
observed, was most likely very narrow. GRB 021206 is among the
strongest bursts ever. Adopting the energy-angle relation we find
an estimated angle of $<1/40$rad or even smaller. Thus, for this
particular burst the required geometry is not unusual. We conclude
that the RHESSI observations  suggest that the prompt emission
results from synchrotron radiation. However, in view of the
comparable levels of polarizations predicted by both the random
field and the homogeneous field scenarios these observations are
insufficient to rule out or confirm either one.

\end{abstract}

\section{Introduction}

It is widely accepted that $\gamma$-ray bursts are produced by the
dissipation of energy in a highly relativistic wind, driven by
gravitational collapse of a (few) solar mass object into a neutron
star or a black hole (see e.g. \cite{fireballs1,fireballs2} for
reviews). As the observed radiation is produced at a large
distance from the collapsing object, key questions remain
unanswered. The nature of the collapsing object ("inner engine")
remains unknown, and the mechanisms responsible for conversion of
gravitational energy into a relativistic wind, as well as for
gamma-ray production, are not well understood.

The recent detection of very high linear polarization during the
prompt \( \gamma  \)-ray emission of GRB 021206 (Coburn \& Boggs
2003; hereafter CB03) has important implications related to these
open questions (Waxman 2003; Lyutikov et al. 2003,  hereafter
LPB03; Granot 2003, hereafter G03,). First, it strongly suggests
that synchrotron emission is the basic mechanism operating in GRBs
(see however, Eichler and Levinson, 2003, who suggest that a large
polarization can arise from Compton scattering of the observed
radiation by cold electrons within a very specific geometrical
construction.). Second, based on the observed very high degree of
polarization \( \Pi =(80\pm20)\% \), several recent papers claim
that a coherent magnetic field is a must (CB03; LPB03), or at
least is strongly favored over a random magnetic field (G03). The
latter conclusion has also led to claims that the high
polarization level measured implies that the magnetic field must
be advected from the {}``inner engine\char`\"{} (LPB03), and that
consequently the relativistic flow must be Poynting flux
dominated. On the other hand, it was suggested \cite{W03} that
the high degree of polarization may also arise for a random
magnetic field, in case that the GRB outflow is a narrow jet, of
opening angle $\theta_j\sim 1/\Gamma$ where $\Gamma$ is the jet
Lorentz factor, provided the jet axis is at an angle
$\sim1/\Gamma$ with respect to the line of sight
\cite{Gruzinov01}.

The main goal of this paper is to critically discuss the
implications to GRB models of the observed high degree of
polarization. In \S~\ref{sec:obs} we discuss the observations and
the constraints they impose on the degree of $\gamma$-ray
polarization. In \S~\ref{sec:jets} we present calculations of
polarization from jets with a uniform magnetic field
(\S~\ref{sec:hom}), and with a random magnetic
field(\S~\ref{sec:ran}). We present more comprehensive calculation
than those of LPB03 and G03, taking into account time dependence,
realistic (no-sharp-edge) jet structure, and using  the specific
electrons' distribution which is relevant to the observation of
RHESSI. We find that the observed polarization may be explained
assuming either a uniform or a random magnetic field
configuration. The constraints imposed on the model in each case
are, however, different. We discuss the implications under the
hypothesis of a uniform field, in particular the issue of whether
a uniform field necessarily favors magnetically dominated out
flow, in \S~\ref{sec:imp1}. Implications under the hypothesis of a
random field, in particular the issue of whether the required
$\sim1/\Gamma$ offset between the jet axis and the line-of-sight
is likely or not, are discussed in \S~\ref{sec:imp2}. Our
conclusions are summarized in \S~\ref{sec:conclusions}.

\section{Discussion of the observations}

\label{sec:obs}

The data analysis of CB03 is based on 12 data points which are
collected over 5sec. Each of these points is a sum of several
independent observations taken at different times.  Thus the data
is some kind of convolution of the polarization over the whole
duration of the burst and should be considered as a \textit{time
averaged data}.

CB03 test two hypothesis with this data. First they test the null
hypothesis of no polarization. This hypothesis is rejected at a
confidence level of \( 5.7\sigma  \). It is clear that the
emission is strongly polarized. Second they estimate the
modulation factor assuming a constant polarization direction and
angle during the whole burst. The best fit to the data is achieved
with \( \Pi =(80\pm 20)\% \).  However, CB03 find that the
probability that $\chi^2$ is greater than the value obtained with
this fit is 5\%.

In checking the second hypothesis, CB03 assumed a time independent
polarization as well as a time independent photon spectrum (given
by the photon spectrum averaged over the GRB duration). The
detected polarization signal arises from a correlation between the
time dependence of scattered photon flux and the angular
orientation of the satellite, which varies on a time scale similar
to the burst duration (the satellite rotation period is comparable
to burst duration). Since both the photon spectrum and the
polarization are likely to vary over this time scale, it is not
clear what the lower limit implied by the observation on the
polarization level is. A robust estimate of this limit will
require a comparison of the observed signal with that obtained
under various assumptions regarding the time dependence of the
spectrum and of the polarization.

Our interpretation of   the available analysis of the observations
is that the time integrated polarization of the prompt \( \gamma
\)-ray emission is of the order of tens of percents, and that the
polarization angle does not vary significantly during the whole
burst.

\section{Synchrotron Polarization from Relativistic Moving Jets}

\label{sec:jets} The level of polarization produced by an ordered
magnetic field that is in the plane of the shock was calculated by
LPB03 and Granot \& Konigel (2003), while G03 calculates the
polarization both from random and from ordered magnetic fields
(see also Sari (1999) and Ghisellini \& Lazzati, (1999)) . Below
we repeat these calculations with some minor changes that are
relevant for the interpretation of RHESSI's observation. We also
add a time dependent calculation that shows the variation of the
polarization with time. The energy range in which the
polarization is measured is 0.15-2Mev. We expect that this arises
from a fast cooling synchrotron emission and it is above both the
synchrotron and the cooling frequencies. The spectral index (\(
F_{\nu }\propto \nu ^{-\alpha } \)) in this power law segment is
\( \alpha \approx 1 \), while the dependence of the intensity on
the magnetic field in the fluid comoving frame is \( I'_{\nu
'}\propto B^{\varepsilon } \), with \( \varepsilon
=(p-2)/2\approx 0 \) and \( p\approx 2-2.5 \) the power index of
the electron's energy distribution. The local emitted
polarization of a synchrotron emission at this power law segment
is \( \Pi_{synch}=(p+2)/(p+10/3)\approx 75\% \) (G03). Averaging
over the emitting area reduces the polarization and in the
following discussion we will estimate the resulting polarization
due to this averaging relative to $\Pi_{synch}$. Namely, unless
otherwise specified the results should be multiplied by this
factor.

We consider two cases: ordered (homogeneous) magnetic field and a
random magnetic field. In both cases the magnetic fields are in
the plane of the shock.

\subsection{Homogeneous Magnetic Field}
\label{sec:hom}

Consider a top-hat jet with an opening angle \( \theta _{j} \)
propagating at Lorentz factor \( \Gamma \)
observed from \(\theta _{obs}\ll \theta _{j} \)\footnote{%
Through the paper we work in spherical co-ordinates with the
origin at the source of the explosion. \( \theta  \) is the angle
with respect to the line between the observer and the origin, and
\( \phi  \) is the azimuthal angel. } with a constant magnetic
field in the plane perpendicular to the motion. The observed
stokes parameters are  weighted averages of the local stokes
parameters at different regions of the shell. The instantaneous
polarization is calculated using the instantaneous observed flux
\( F_{\nu }(y,T)\propto (1+y)^{-(3+\alpha )} \) as the weights,
where \( y \equiv (\Gamma \theta )^{2} \) and \( T \) is the
observer time. The time integrated polarization is calculated
using the fluences as weights: \( \int ^{\infty }_{0}F_{\nu
}(y,T)dT\propto (1+y)^{-(2+\alpha)} \).

For \( \varepsilon =0 \) the time integrated stokes parameters
(note that \( V=0 \) as the polarization is linear) and
polarization are given by:
\begin{equation} \label{Eq QU ordered} \frac{\left\{ \begin{array}{c}Q \\U \\
\end{array}\right\}}{I}=\Pi _{synch} \frac{\int _{0}^{2\pi }\int _{0}^{\infty }(1+y)^{-(2+\alpha
)}\left\{ \begin{array}{c} \cos(2\theta _{p}) \\ \sin(2\theta _{p}) \\
\end{array}\right\}dyd\phi }{\int _{0}^{2\pi }\int _{0}^{\infty
}(1+y)
 ^{-(2+\alpha )}dyd\phi } ,
\end{equation}

and
\begin{equation} \label{Eq Pi} \Pi
=\frac{\sqrt{U^{2}+Q^{2}}}{I},
\end{equation}
where \( \theta _{p}=\phi +\arctan(\frac{1-y}{1+y}\cot\phi ) \)
\cite{GK03}. This result is similar to the result of LPB03 apart
for the fact that LPB03 calculate the polarization for a slow
cooling emission with \( \nu _{m}<\nu <\nu _{c} \), (with
different values of $\alpha$ and $\epsilon$) which is not the
relevant case for the prompt $\gamma$-ray emission that was
observed by RHESSI. For \( \alpha =1 \) Eqs. \ref{Eq QU
ordered}-\ref{Eq Pi} yield a polarization level of \( \Pi /\Pi
_{synch}\approx 60\% \). I.e. 60\% of the maximal synchrotron
polarization, or an overall polarization of $\sim 45\%$  (taking
the exact values of $\alpha$ and $\epsilon$ for $p=2.5$ results
in an overall polarization of $\sim 50\%$).

The time dependent polarization depends on the the temporal
properties of the emitting area. We consider the simplest case of
a thin shell radiating at a constant luminosity between \( R_{1}
\) and \( R_{2} \). The first photon arrives to the observer at \(
T_{1}=R_{1}/(2c\Gamma ^{2}) \). At this time the observer sees
only photons from the center (\( y=0 \)), thus the initial
polarization is \( 100\%. \) Later the observer receives photons
both from the center and from higher angles and the polarization
drops to an asymptotic value which is reached when photons from
the whole shell reach the observer simultaneously. Finally,
during the decay of the observed pulse the observer receives only
photons from \( y>0 \) and the polarization level drops. More
precisely at time \( T \) the observer receive photons from \(
y_{min}<y<y_{max} \) where:
\begin{equation}
\label{Eq Yminmax}
\begin{array}{c}
y_{min}(\widetilde{T})=max(0,\widetilde{T}/\widetilde{T_{2}}-1)\\
y_{max}(\widetilde{T})=\widetilde{T}-1
\end{array},
\end{equation}
where \( \widetilde{T}=T/T_{1} \) and \(
\widetilde{T_{2}}=R_{2}/R_{1} \). Now the polarization at \(
\widetilde{T} \) is given by:
\begin{equation}
\label{Eq QU(T) ord}
\frac{\left\{ \begin{array}{c}Q \\U \\
\end{array}\right\}}{I}(\widetilde{T})=\frac{\int_{0}^{2\pi}\int
_{y_{min}}^{y_{max}}(1+y)^{-(3+\alpha )}\left\{ \begin{array}{c}
\cos(2\theta _{p}) \\ \sin(2\theta _{p})\\
\end{array}\right\}dyd\phi }{\int _{0}^{2\pi
}\int_{y_{min}}^{y_{max}}(1+y)^{-(3+\alpha )}dyd\phi }
\Pi_{synch}.
\end{equation}

\begin{figure}
{\par\centering
\resizebox*{0.9\columnwidth}{0.3\textheight}{\includegraphics{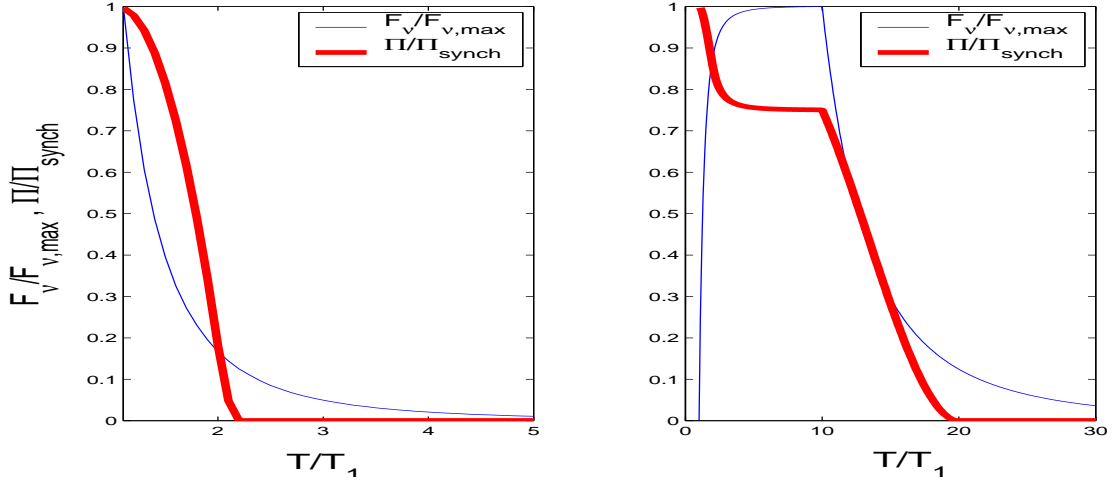}}
\par} \caption{The instantaneous polarization ({\it thick line})
and normalized flux ({\it thin line}) from a thin shell
propagating at a Lorentz factor $\Gamma$ and radiating constantly
between the radii $R_1$ and $R_2$ under the hypothesis of uniform
and constant magnetic field. The x axis is the observer time $T$
in units of the angular time at $R_1$: $T_1=R_1/(2c\Gamma^2)$. In
the left panel $R_2=1.01R_1$, in the right panel $R_2=10R_1$ and
we can see the gradual increase in flux and the corresponding
decrease in the polarization that reaches an asymptotic value
when the whole shell is observed.}
\end{figure}

Figure 1 depicts the time dependent polarization from a shell
emitting between $R_1$ and $R_2$. When \( (R_{2}-R_{1})/R_{1}\ll 1
\) (the radial time ,\( (R_{2}-R_{1})/2c\Gamma ^
 {2} \), is much
shorter than the angular time (\( R_{1}/(2c\Gamma ^{2}) \)), left
panel, the polarization level drops from \( 100\% \) to a few
percent within one angular time. When the radial time is much
larger than the angular time , right panel, the polarization level
drops from \( 100\% \) to an asymptotic value until the peak of
the pulse. As the pulse begins to decay the polarization drops to
a few percent within one angular time of $R_2$.

\subsection{Polarization from a Random Magnetic Field}
\label{sec:ran}

Waxman (2003) has suggested that a high polarization level can be
obtained when a narrow jet that is observed from the edge even if
the magnetic field is random. We consider here a random magnetic
field that remains planner in the plane of the shock. For a three
dimensional random magnetic field the polarization essentially
vanishes. For \(\epsilon = (p-2)/2 \approx 0 \) the degree of
observed polarization of the emission emitted from a small region
at angle $y$ is: \( \Pi (y)/\Pi _{synch}=min(y,1/y) \). The
overall time integrated stokes parameters are:
\begin{equation}
\label{Eq QU rand} \frac{\left\{ \begin{array}{c}Q \\U \\
\end{array}\right\}}{I}=\Pi _{synch}\frac{\int _{0}^{2\pi
}\int_{0}^{\infty}P'_{\nu',m}(1+y)^{-(2+\alpha)}\min(y,1/y)\left\{
\begin{array}{c}
\cos(2\phi) \\ \sin(2\phi)\\
\end{array}\right\}dyd\phi }{\int _{0}^{2\pi }\int_{0}^{\infty }P'_{\nu
',m}(1+y)^{-(2+\alpha )}dyd\phi},
\end{equation}
where \( P'_{\nu ',m}=P'_{\nu ',m}(y,\phi ) \) is the emitted
power at the synchrotron frequency in the fluid rest frame. For a
top-hat jet with sharp edges \( P'_{\nu ',m} \) is constant for
any \( y \) and \( \phi \) within the jet and zero otherwise. For
a structured jet \( P'_{\nu ',m} \) depends on the angle from the
jet axis.

The maximal polarization is observed when we see the edge of the
jet. The probability to see the edge of a top-hat jet with sharp
edges and an opening angle \( \theta _{j} \Gamma \gg 1 \) is
negligible. On the other hand a jet with \( \theta _{j} \Gamma \ll
1 \) is not expected. Thus the only physical cases in which we
can expect a large polarization are \( 1 \mla \theta _{j}\Gamma
<{\textrm{a few }} \).

\begin{figure}
\resizebox*{0.45\columnwidth}{0.3\textheight}{\includegraphics{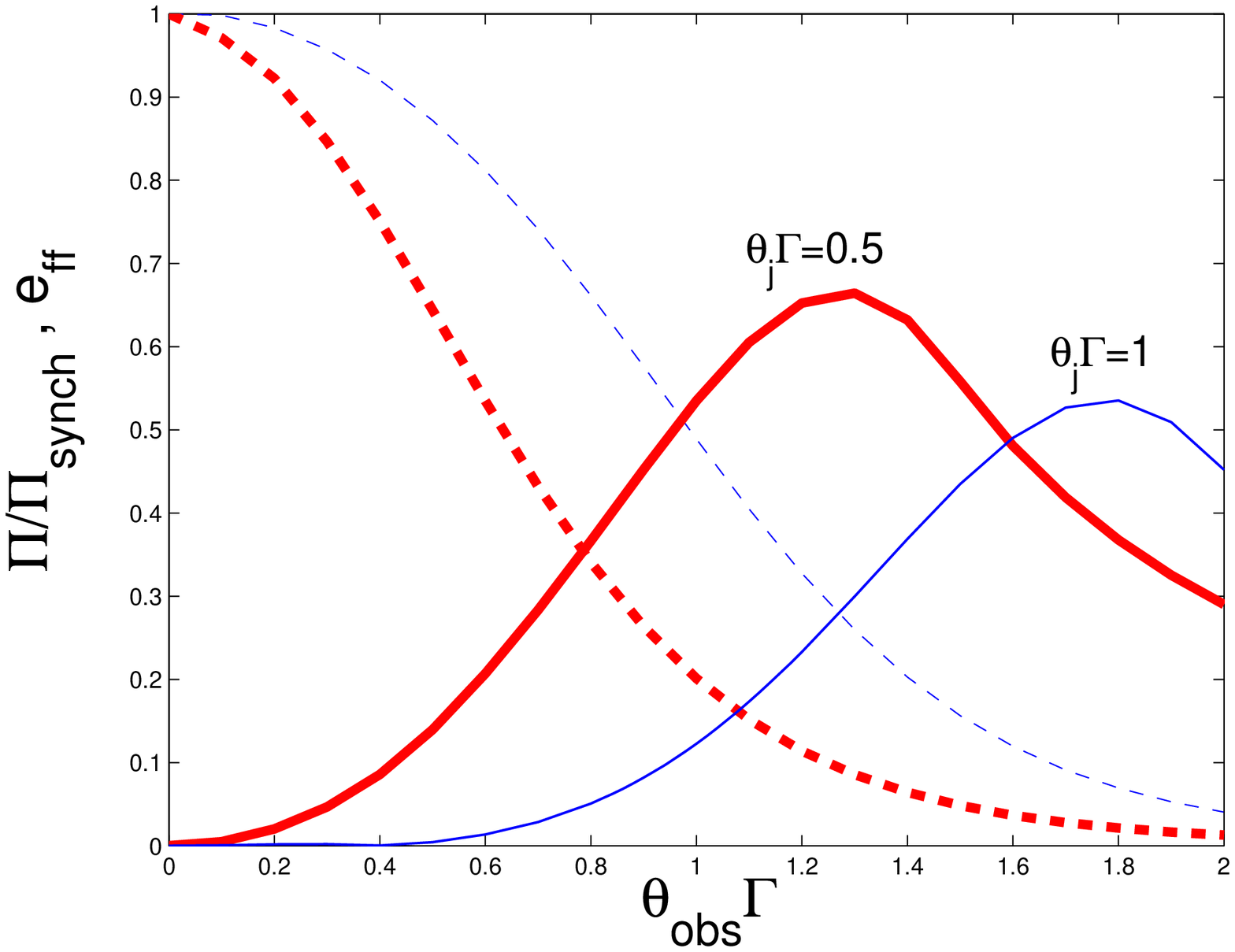}}
\resizebox*{0.45\columnwidth}{0.3\textheight}{\includegraphics{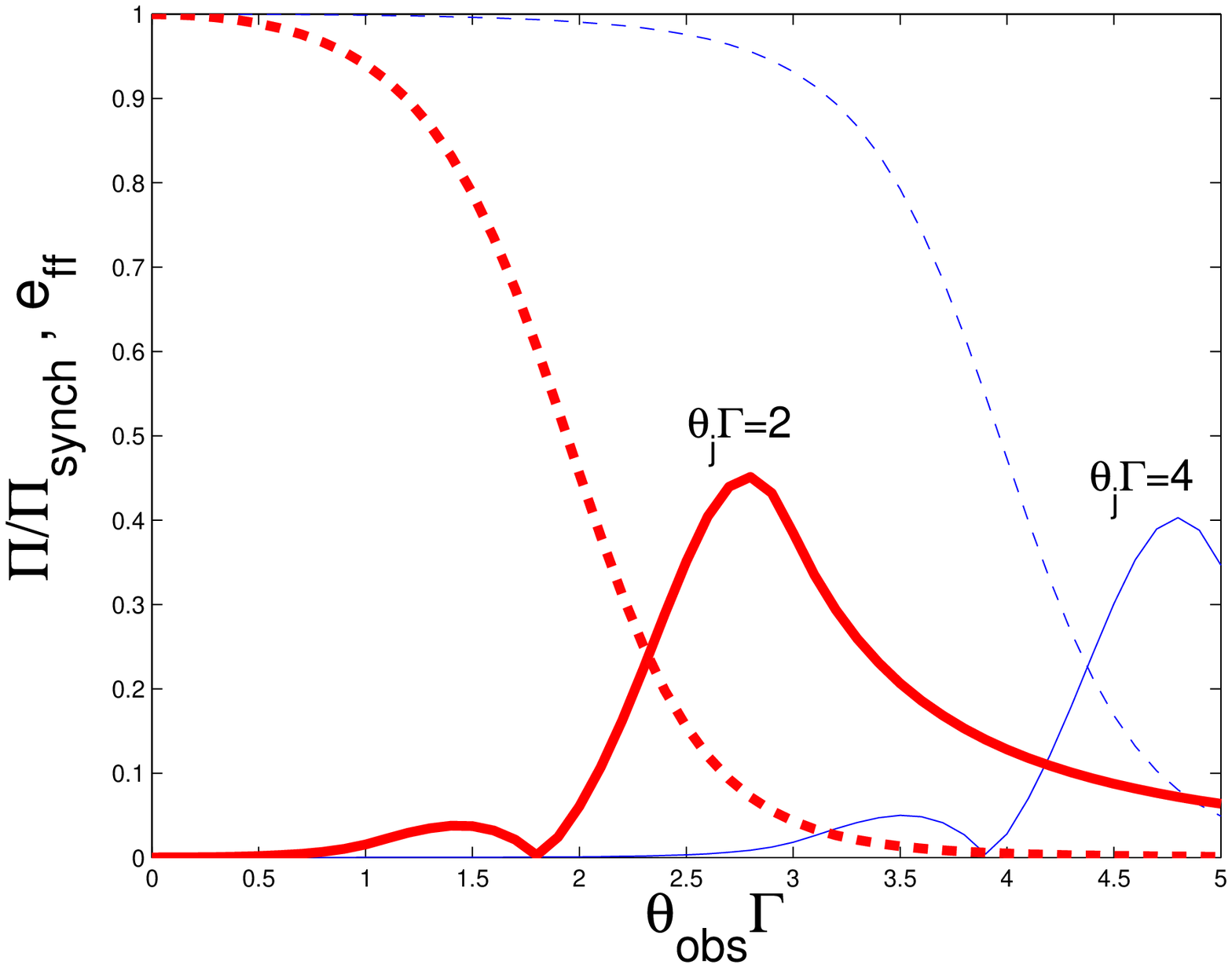}}
\caption{The time integrated polarization ({\it solid line}) and
the efficiency ({\it dashed line}) as a function of $\theta_{obs}
\Gamma$ for four different values of $\theta_j$ for a random
magnetic field.}
\end{figure}

Figure 2 depicts the time integrated polarization and the
efficiency from sharp edged jets with different opening angles as
a function of the the angle between the jet axis and the line of
sight, \( \theta _{obs} \). The efficiency, \( e_{ff} \) is
defined to be the ratio between the observed fluence at \(
\theta_{obs} \) and the maximal possible observed fluence at \(
\theta _{obs}=0 \). In all these cases the polarization is peaked
above 40\%, however the efficiency decrease sharply as the
polarization increase. Thus the probability to see high
polarization grows when \( \theta _{j} \) decrease. The
probability that \( \theta _{obs} \) is such that the polarization
is larger than \( 30\% \) ($\cdot \Pi_{synch}$) while \( e_{ff}>0.1 \) is
0.68, 0.41, 0.2 \& 0.08 for \( \theta _{j} \Gamma =0.5,1,2,4 \)
respectively. In reality this probability will be smaller, as the
chance to observe a burst increases with its observed flux.

\begin{figure}
\resizebox*{0.45\columnwidth}{0.3\textheight}{\includegraphics{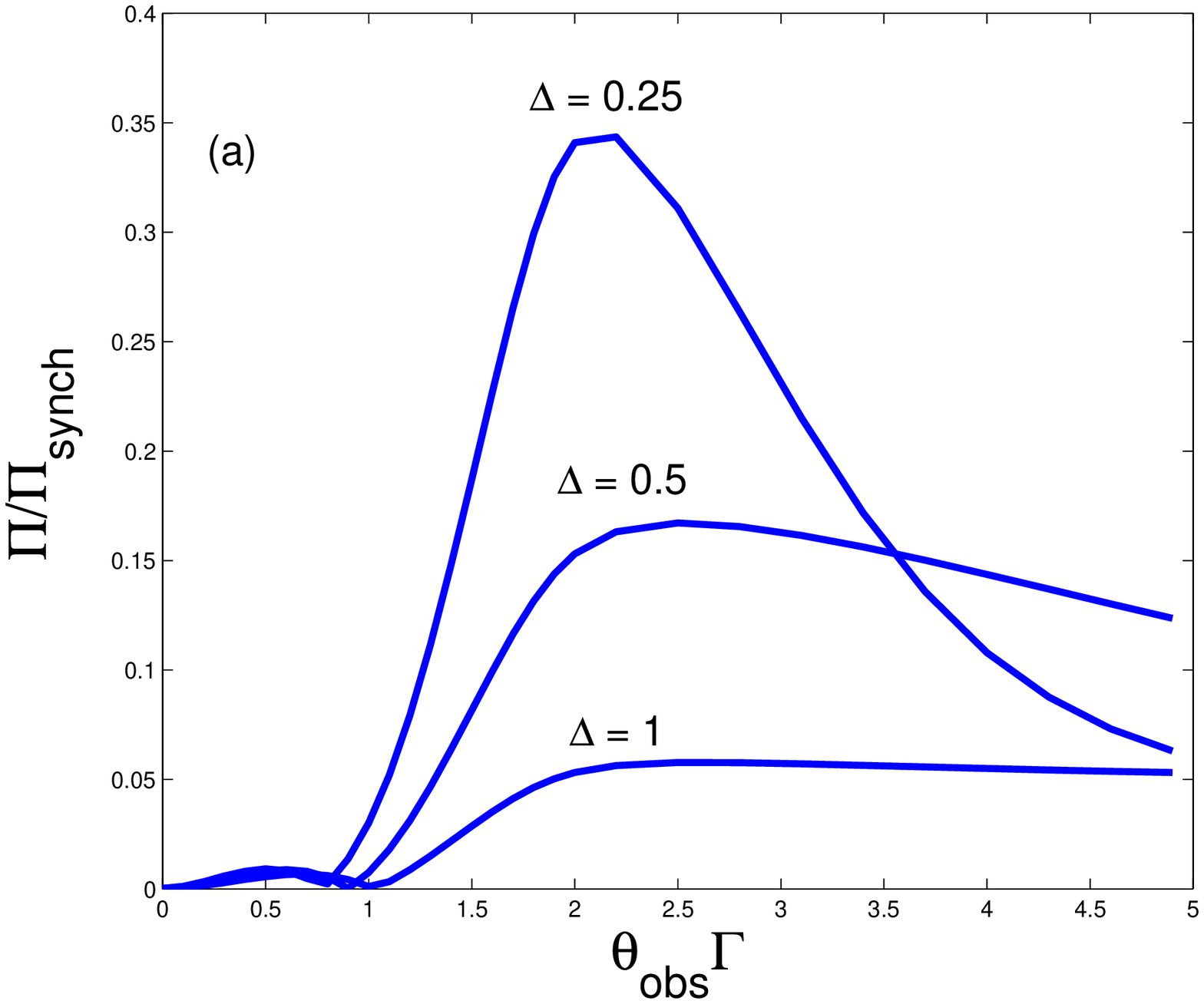}}
\resizebox*{0.45\columnwidth}{0.3\textheight}{\includegraphics{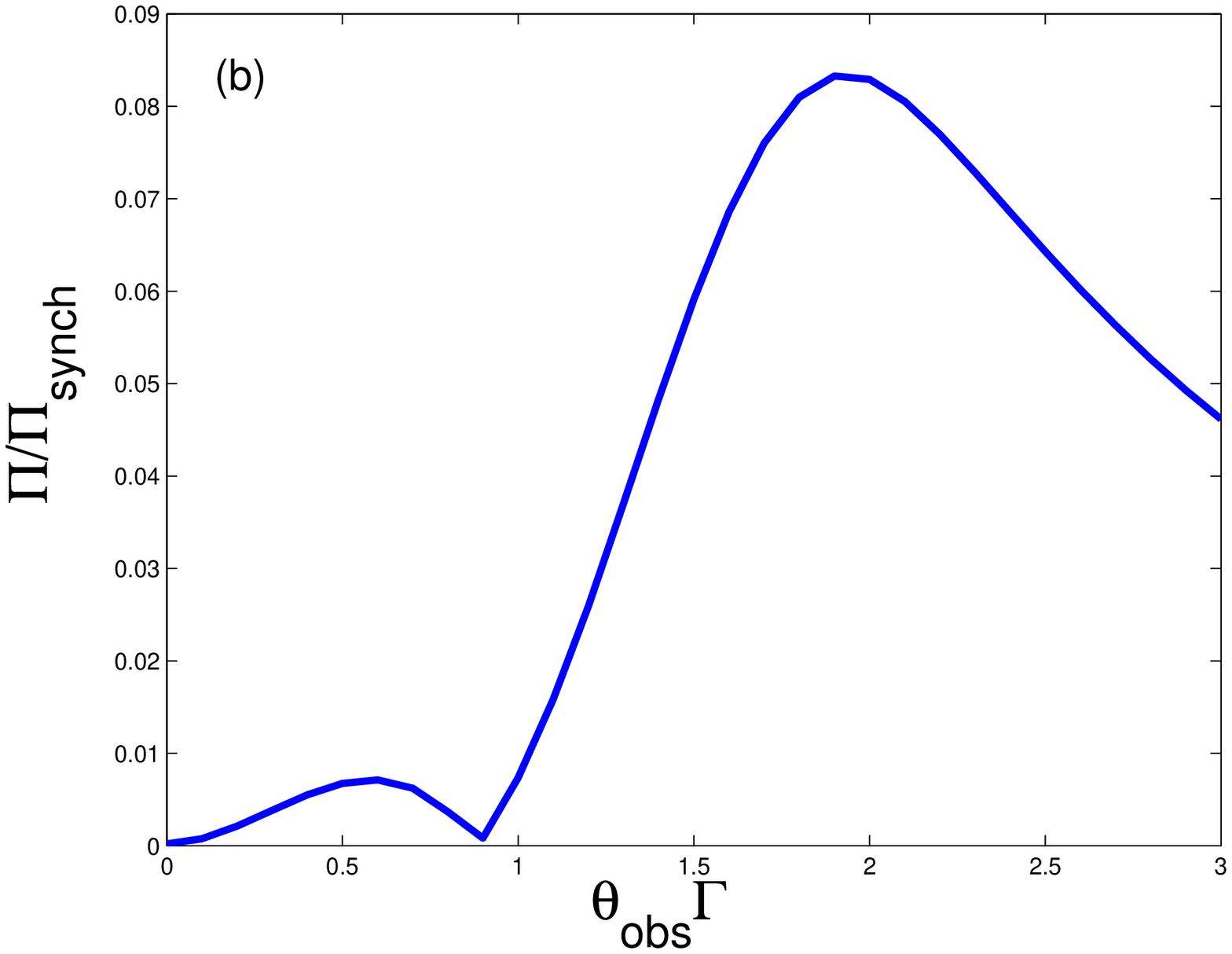}}
\caption{The time integrated polarization of a jet with
exponantial wings ({\it left panel}, according to Eq. \ref{Eq exp
wing}) and power-law
wings ({\it right panel}, according to Eq. \ref{Eq powerlaw wing}),
as a function of
$\theta_{obs}$. The size of the core of the jet (over which the
emissivity is constant) is $1/\Gamma$. }
\end{figure}

Figure 3a depict the polarization produced by a jet with
exponential wings. The emitted power of the jet as a function of
the angle with respect to the jet axis, \( \psi  \) is:
\begin{equation}
\label{Eq exp wing}
P'_{\nu',m}=\left\{ \begin{array}{c}P'_{\nu ',m}(\psi =0)\quad
\psi \Gamma <1\\P'_{\nu ',m}(\psi =0)\, \exp[(\psi \Gamma
-1)/\Delta
 ]\quad \psi \Gamma >1\end{array}\right. ,
\end{equation}
where \( \Delta  \) is the characteristic angle (in units of \(
1/\Gamma \)) over which the emission decays. According to Fig.3a
it is clear that in order to obtain high polarization the wings of
the jet must be sharp (\(\Delta <0.25 \)). An alternative possible
wing is a power law wing:
\begin{equation}
\label{Eq powerlaw wing}
P'_{\nu ',m}=\left\{ \begin{array}{c} P'_{\nu ',m}(\psi =0)\quad
\psi <\psi _{c}\\P'_{\nu ',m}(\psi =0)\, (\psi /\psi
_{c})^{-\delta }\quad \psi >\psi _{c}
\end{array}\right.
\end{equation}
The case of \( \delta =2 \) is especially interesting as it is the
jet structure of the universal standard jet \cite{Rossi02}\footnote{%
Note that in the universal standard jet this is the profile of
the energy and not of the emitted power. However if the radiated
efficiency is constant both profiles are similar.}. Due to
similar considerations as above \( \psi _{c}\Gamma \mga 1 \).
Figure 3b shows that \( \psi _{c}\Gamma =1 \) can not produce a
large polarization.

Similarly to the ordered magnetic field case, the time dependent
polarization depends on the the temporal properties of the
emitting area. Again we consider the simplest case of a thin shell
which radiates at constant luminosity between \( R_{1} \) and \(
R_{2} \). The calculation is also similar to the case of ordered
magnetic field:

\begin{equation}
\label{Eq QU(T) rand} \frac{\left\{ \begin{array}{c}Q \\U \\
\end{array}\right\}}{I}(\widetilde{T})=\Pi_{synch}\frac{\int
_{0}^{2\pi }\int _{y_{min}}^{y_{max}}P'_{\nu',m}(1+y)^{-(3+\alpha
)}\min(y,1/y)\left\{
\begin{array}{c}
\cos(2\phi) \\ \sin(2\phi)\\
\end{array}\right\}dyd\phi }{\int_{0}^{2\pi }\int
_{y_{min}}^{y_{max}}P'_{\nu',m}(1+y)^{-(3+\alpha )}dyd\phi },
\end{equation}
\begin{figure}
\resizebox*{0.9\columnwidth}{0.3\textheight}{\includegraphics{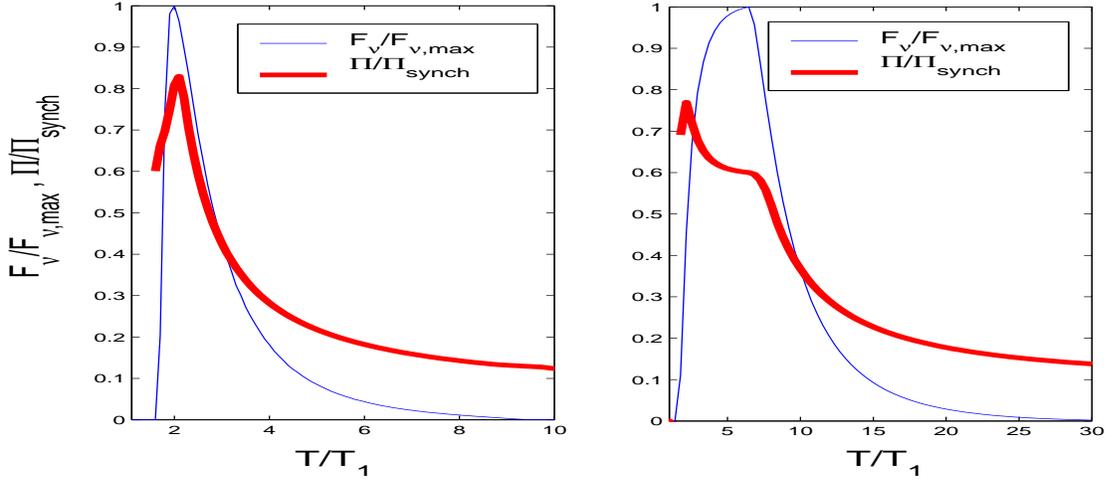}}
\caption{The instantaneous polarization ({\it thick line}) and
normalized flux ({\it thin line}) from a thin shell which
propagates at a Lorentz factor $\Gamma$ and radiates constantly
between the radii $R_1$ and $R_2$ under the hypothesis of {\it random}
magnetic field in the plane of the shock. The x axis is the observer time $T$
in units of the angular time at $R_1$: $T_1=R_1/(2c\Gamma^2)$. In
the left panel $R_2=1.1R_1$, in the right panel $R_2=4R_1$.}
\end{figure}
where $y_{min,max}$ are defined in Eq. \ref{Eq Yminmax}. Figure 4
depicts the time dependent polarization for a top-hat jet with \(
\theta _{j}\Gamma =1 \), observed from \( \theta_{obs}\Gamma =1.8
\). When the radial time is much shorter than the angular time,
left panel, the polarization level follows the flux of the pulse
and at its peak it can reach more then 80\% ($\cdot \Pi_{synch}$). When the
radial time is much larger than the angular time , right panel,
the polarization level rises at first and than it drops to an
asymptotic level till the peak of the pulse. As the pulse starts
to decay, the polarization resumes its decay as well.

So far we have assumed that the Lorentz factor, $\Gamma$ does not
vary during the burst. Large variation in the Lorentz factor will
induce variation in the averaging regions and will reduce the
observed polarization by a factor of a few (G03), making it too
low. However, if the modulated wind is a constant energy flow
(which is favored due to other reasons \cite{NP02}) then the
spread of the Lorentz factor of the emitting regions is small,
and no significant reduction in the level of polarization is
expected. Furthermore, even if there are variations in the
Lorentz factor we would expect that the variations in the flux
weighted Loretnz factor will be rather small.

\section{Implications for a uniform magnetic field}
\label{sec:imp1} We discuss in this section the implications of
the high level of polarization, under the hypothesis of a uniform
magnetic field configuration. We first show
 (\S~\ref{sec:causal})
that "uniformity" means "uniform field over a causally connected
region" of the outflow, and then discuss the implications
(\S~\ref{sec:transport},\S~\ref{sec:acceleration}).

\subsection{Coherence scale and causal connection}
\label{sec:causal} A distant observer receives radiation from a
conical section of the outflow, of opening angle $\sim1/\Gamma$
where $\Gamma$ is the Lorentz factor of the out flowing plasma.
The transverse size of the observed region is therefore $\sim
R/\Gamma$, where $R$ is the radius at which radiation is
produced. This length scale is comparable to the size of a
causally connected region within the flow, as demonstrated by the
following simple argument.

Consider a signal emitted at radius $R_0$ and propagating at a
speed $\beta_s c$, where $c$ is the speed of light, in the plasma
rest frame. By the time the plasma expands to radius $2R_0$, the
time that elapsed in the plasma frame is $R_0/\Gamma c$, and hence
the distance
 travelled by the signal is $\beta_s R_0/\Gamma$, i.e.
the transversal size over which the signal has been "communicated"
is $\beta _{s}R_0/\Gamma$. Since this transverse size is linear
in $R_0$ for doubling of $R$, it follows that distance travelled
by the signal grows logarithmically with $R$, and is of order \(
\beta _{s}[R/\Gamma ]\log (R/R_{0}) \). This result is independent
of the nature of the flow, and holds regardless of whether the
energy density in the outflow is carried mainly by magnetic field
or particles. It is simply a results of causality. Note that
LPB03 base their criticism of the hydrodynamic scenario on a
different calculation of the size of the causally connected
regions. They consider signals that propagate at a constant speed
in the observer frame (and ignore the fact that a hydorodynamic
wave is dragged along the flow)  and obtain a different result.
Our result implies that under the "uniform field" scenario, the
field is required to be uniform over causally connected regions
both in hydrodynamic flows and in Poynting flux dominated flows.

\subsection{Transport from the source}

\label{sec:transport}

There are two qualitatively different scenarios for the
production of the strong magnetic field implied to exist in the
synchrotron emitting region. The field may be produced near the
compact object driving the wind (as originally suggested by
\cite{Usov94}), and then carried by the wind to large radii where
radiation is emitted. Alternatively, it may be produced (or
strongly amplified) in the radiation emitting region itself. The
common mechanism assumed to be responsible for the latter
scenario is magnetic field amplification by plasma instabilities
in collisionless shocks. The magnetic field produced by the
latter process is naturally created on small (skin-depth) scales
(e.g. \cite{GW99,ML99}), and there is no well understood process
that increases the field coherence scale to a size comparable to
the that of a causally connected region within the outflow.
Nevertheless, afterglow observations strongly suggest that the
field is created by the collisionless shocks, and that its
coherence length does grow to the size of causally connected
regions \cite{GW99}. However, if the field is produced by this
process, we would expect its direction to vary with time on a time
scale similar to that of the time scale for $\gamma$-ray flux
variations. In this case the resulting time integrated
polarization would be similar to the one produced by a random
field  for a variable burst as GRB 021206. In order to obtain the
large time averaged polarization observed by CB03, under the
hypothesis of uniform field, the field must be coherent over
causally connected regions and {\it time independent}.

The most natural assumption that would lead to a coherent time
independent field is that the field is produced by the compact
object, and then carried by the wind to large distances. LPB have
shown that in the case where the wind is cylindrically symmetric
and dominated by Poynting flux, assuming that the inertia of the
plasma is negligible ("force free" approximation), the field is
dominated at large distances by a coherent toroidal component,
$B_\phi$. In this case, large polarization will arise if the
expanding plasma is observed at an angle $>1/\Gamma$ with respect
to the symmetry axis. Note that this torodial geometry will keep
the direction of the magnetic field constant in time, as required
under the uniform field hypothesis,  even if the wind is not
steady.

However, coherent toroidal field may also be obtained in the case
where the energy density is not dominated by Poynting flux. In the
case where the wind energy density is dominated by the plasma
particles, and where the conductivity is high so that the electric
field in the plasma rest frame
($\vec{E'}=\vec{E}+c^{-1}\vec{v}\times\vec{B}$) is vanishingly
small (the MHD approximation), Maxwell's eqs. give, for a steady
wind, $\nabla\times(\vec{v}\times\vec{B})=0$, which imply
$B_\phi\propto R^{-1}$ for uniform radial velocity. Thus, in this
case too a coherent toroidal field dominates at large radii, with
energy flux that is radius independent. In order to account for
the observed $\gamma$-ray emission it is sufficient that the
magnetic field carries $<1\%$ of the wind luminosity.

\subsection{Particle acceleration}

\label{sec:acceleration}

Regardless of whether or not the wind energy density is dominated
by Poynting flux, the observed $\gamma$-rays are produced by the
dissipation of wind energy, which converts either magnetic field
energy or plasma kinetic energy to high energy particles
(electrons), which then radiate away the dissipated energy. In the
kinetic energy dissipation scenario, particles are assumed to be
accelerated in collisionless shocks. In this case, non uniform
field components must be produced in the shocks, since particles
of a wide range of energies and Larmor radii must be efficiently
scattered across the shock. It may therefore appear that particle
acceleration and uniform magnetic field are mutually exclusive.

A similar problem arises also in the case of a Poynting flux
dominated flow. LPB do not present a model for the dissipation
 of
magnetic field energy and particle acceleration. However, it
appears to be unlikely that such dissipation, which converts a
significant fraction of field energy to particle energy, may arise
without exciting waves of a wide range of wavelengths and wave
vector orientations resulting in a large fraction of the energy
density converted to a random field component. \cite{Smolsky00}
propose a specific mechanism for magnetic field energy
dissipation. However, since their calculations are one
dimensional, it is difficult to draw conclusions regarding the
resulting polarization.

A possible solution to this apparent contradiction between uniform
field and particle acceleration is that a non negligible fraction
of the magnetic field remains coherent (and responsible for the
polarization) while a random component with non negligible
fraction of the energy density is produced via dissipation and is
responsible for particle acceleration.  However, in this
configuration the radiation from the random component of the
magnetic field would reduce the polarization level. A second
solution may be a separation of the acceleration region and the
emission regions. However, no detailed physical solution of this
type has been presented so far.

\section{Implications for a random magnetic field}

\label{sec:imp2}

An off-axis observation can produce, in the case of a random
magnetic field configuration, up to 60\% ($\cdot \Pi_{synch}$)
time integrated polarization (although a value of
30\%($\cdot \Pi_{synch}$) is much more probable) and
an instantaneous polarization which can peak around 80\% (see
\S~\ref{sec:ran}). These results are quite similar to the
polarization level that is produced by a uniform time independent
magnetic field. While the main difficulty in the uniform time
independent field scenario is the production of the
  field
(\S~\ref{sec:transport}) and the particle acceleration
(\S~\ref{sec:acceleration}) within such a field, the physical
conditions which may produce a random magnetic field are much
more relaxed. However, the main drawback of the random magnetic
field scenario is that not all the observers see the same level
of polarization. In other words, a high degree of polarization
during the prompt emission is possible only if $\theta_{obs}
\approx \theta_j \approx 1/\Gamma \approx 0.01$rad. Such
orientation may appear unlikely. However, we argue below that
this is not necessarily the case in GRB 021206.

The opening angle of GRB 021206 can be estimated using the
constant total energy of GRBs \cite{F01,PK01}. The observed
fluence of GRB 021206 was $1.6 \cdot 10^{-4}ergs/cm^2$ at the
energy range of 25-100Kev \cite{Hurley02}. This puts GRB 021206
as one of the most powerfull bursts, and the most powerful one (a
factor of 2-3 above GRB990123) after correcting for the fact that
it was observed only in a narrow band (compared to the wide BATSE
band of 20-2000keV). We correct for this factor using the data
for GRB 020813 that was also observed by the IPN at the same
waveband. This burst that was at z=1.25 had an average total
(beaming corrected) energy of $1.2 \cdot 10^{51} \rm ergs$ (which
is the typical energy of GRBs), a fluence lower by a factor of
four (compared to GRB 021206) and had an opening angle of $3^o$
(see Tables 1 and 2 of Bloom et al. (2003)). Assuming,
conservatively, that the redshift of GRB 021206 is comparable we
find $\theta_j \approx 1.5^o \approx 1/40$rad, putting GRB 021206
as the narrowest jet seen so far (a factor of 2 below the
narrowest jets found by Bloom et al. (2003)). Note that the
correction for varying the redshift in the range 1-2 is small.

This simple estimate of the opening angle yields $\theta_j$ which
is a factor 2-3 from the typical (inverse) Lorentz factors seen in
GRBs. We can refine the estimate.
 As we have shown above (\S~\ref{sec:ran}) in order to obtain
high polarization the jet must be seen from the edge. This
implies that we detected only a fraction $e_{ff}$ of the full
flux of the burst. With $e_{ff} \approx 0.1$ this will introduce
an additional correction of $\sim 1/3$ to the opening angle,
putting the overall angle around $1/120$rad. Thus, we find that
the condition $\theta_j \approx 1/\Gamma$ is very reasonable for
this particular burst. We also find that as GRB 021206 is clearly
atypical, as seen from its remarkably high fluence, one does not
have to worry too much about the issue of whether the observation
of the jet from the edge is generic or not.

\section{Conclusions}

\label{sec:conclusions}

The recent detection of very high linear polarization during the
prompt \( \gamma  \)-ray emission of GRB 021206 (CB03) suggests
strongly that the $\gamma$-rays are produced by synchrotron
emission of relativistic particles in strong magnetic field. The
intrinsic linear polarization of such emission can reach $75\%$
(depending weakly on the details of the electron's energy
distribution) and it is perpendicular to the magnetic field and
to the photon's momentum. In GRBs the emission arises from a
relativistic jet and the observed polarization should be averaged
over the emitting region taking into account relativistic effects
that rotate the propagation vector and limit the observing angle
to a region of a size $\Gamma^{-1}$. This averaging essentially
reduces the polarization from its local maximal value. The
relativistic rotation of the polarization vector leads to a
decrease in the overall polarization even in cases that the whole
emitting region is emitting locally fully polarized radiation.

We have calculated the polarization level of synchrotron radiation
from relativistic jets with either constant (uniform in space and
time independent) or random (but in the plane perpendicular to the
motion) magnetic fields. For a homogeneous field one can obtain a
maximal time integrated polarization of 60-65\% of the maximal
synchrotron polarization i.e. an overall polarization of 45-50\%.
It is interesting to note that even for a constant magnetic field
the instantaneous polarization level (but not the direction)
varies with time and the extend of the temporal variability
depend on the ratio of the intrinsic pulse duration to the angular
spreading time $R/\Gamma^2$ across the jet.

The natural way to produce a time independent homogeneous magnetic
field across the jet is to transport the magnetic field from the
source (see \S \ref{sec:transport}). The magnetic field required
to produce the observed synchrotron emission is rather strong and
it implies both a $\sim 10^{12}$G magnetic fields at the source
and a transport of the field $\propto R^{-1}$. As only the
torodial component of the field can be transported as $R^{-1}$ we
must be looking at the system at least $1/\Gamma$ away from the
symmetry axis. The magnitudes of the field involved are large but
they  do not necessarily imply that most of the wind energy has
to be carried as a Poynting flux. Moreover, the constrain of a
homogeneous magnetic field over the observable region
($1/\Gamma$) does not rule out a baryons dominated flow. Thus
even if the observed polarization results from a homogeneous
magnetic field, it is not enough to distinguish between a
Poynting flux and non-Poynting flux dominated flows. There is a
serious difficulty, within the homogeneous field model for
accounting for particle acceleration both for Poynting flux and
non-Poynting flux domination as the standard Fermi acceleration
processes require a diffusion in a random magnetic field (see \S
\ref{sec:acceleration}). Even within an alternative acceleration
model based on field reconnection rather than on Fermi process
this can be resolved only by separating the acceleration and
emission regions, or by producing a co-existing random and
ordered magnetic fields.

The maximal time integrated polarization from a random magnetic
field depends critically on the viewing angle, $\theta_{obs}$. It
increases with $\theta_{obs}$, reaches a maximum around
$\theta_{obs}=1/\Gamma + \theta_j$ and decreases to zero.
 However as the polarization increases when the observer moves
outwards, the observer views a smaller fraction of the jet and
$e_{ff}$ decreases. With reasonable parameters one can obtain a
polarization of 40\% of the maximal synchrotron polarization i.e.
an overall polarization of 30\%  and $e_{ff} \approx 0.1$. Thus,
somewhat surprisingly, a random magnetic field  produces almost
as strong polarization as the homogeneous one. This arises
because in the homogeneous case the
center contributes of the higher polarization while the
peripheral regions at angle of $1/\Gamma$ reduce it, while in the
random case the peripheral regions contribute to the high
polarization while the center reduces it.

The main challenge for the random field scenario is the jet
opening angle and orientation (see \S \ref{sec:imp2}). However,
for this particular burst we find that the required geometry is
likely if we adopt the energy-angle relation (Frail et al. 2001).
The probability to observe the edge of the jet depends on
$\theta_j \Gamma$. It will be significant only if $\theta_j
\Gamma \approx 1$ or at least it is not significantly larger than
unity. This last point may indicate that the situation of
observing large polarization from a jet with random magnetic
fields is not generic. However, we have shown that GRB 021206 is
not generic. Its fluence put it as one of the strongest GRBs
observed ever and according to the constant energy reservoir
\cite{F01} interpretation this implies that $\theta_j$ was one of
the narrowest. Conservative estimates that put it at $z \approx
1.25$ suggest $\theta_j \approx 1/40$rad and inclusion of the
efficiency factor yield $\theta_j \approx 1/120$rad, values that
almost comparable to various estimates of $1/\Gamma$. We have also
found that the jet needs to be reasonably sharp, but not
unrealistically so.

Over all we find that both cases are incompatible with the
canonical value of 80\% estimated by CB03. However, the data has
been fitted with only two models:  (i) no polarization that was
rejected and (ii) a constant polarization of 80\%, which  also
does not provide a perfect fit to the data. On the other hand
both the homogeneous field and the random field scenarios suggest
that the polarization degree (and possibly the orientation) vary
with time. We conclude that the current data indicates that the
polarization level was high, suggesting synchrotron that has
intrinsic high polarization as the origin of the prompt emission.
However,  in view of the comparable levels of polarizations
predicted by both the random field and the homogeneous field
scenarios the current data is insufficient to rule out or confirm
either one.

We thank J. Granot, D. Guetta and A. Lazar and for helpful
discussions and comments.

\end{document}